\documentclass[aps,pra,showpacs,floatfix,twocolumn,superscriptaddress]{revtex4}
\usepackage{graphicx}
\usepackage{bm,amsmath}
\usepackage{dcolumn}
\usepackage{mathrsfs}
\usepackage{amssymb}
\usepackage{cases}
\usepackage{color}
\usepackage{ulem}
\usepackage{indentfirst}
\newcommand{\be}{\begin{equation}}
\newcommand{\ee}{\end{equation}}
\newcommand{\bea}{\begin{eqnarray}}
\newcommand{\eea}{\end{eqnarray}}
\newcommand{\la}{\langle}
\newcommand{\ra}{\rangle}
\begin{document}
\title{ Trimer superfluid and supersolid on two-dimensional optical lattices}
\author{Wanzhou Zhang}
\thanks{Corresponding author: zhangwanzhou@tyut.edu.cn}
\affiliation{College of Physics and Optoelectronics, Taiyuan University of Technology Shanxi 030024, China}
\author{Yuan Yang}
\affiliation{College of Materials Science and Engineering, Taiyuan University of Technology, Taiyuan 030024}

\author{Lijuan Guo}
\affiliation{College of Physics and Optoelectronics, Taiyuan University of Technology Shanxi 030024, China}
\author{Chengxiang Ding}
\affiliation{Department of Applied Physics, Anhui University of Technology, Maanshan 243002, China}

\author{Tony C Scott}
\affiliation{College of Physics and Optoelectronics, Taiyuan University of Technology Shanxi 030024, China}

\date{\today}
\begin{abstract}
By the photoassociation method, the trimer superfluid phase disappears in the one dimensional
 state-dependent optical lattice if the ratio of the three-body interaction $W$ to the trimer tunneling $J$
 is kept at $W/J=12$ [Phys Rev A. {\bf 90}, 033622(2014)].
To search for a trimer superfluid and trimer supersolid, we
load the cold atom into  two-dimensional lattices,
whose coordinate number $z$ and kinetic energy $-zJ$ are respectively larger and lower
 than those of a one dimensional lattice.
Herein, we study the  Bose-Hubbard model,
which has an additional trimer tunneling term, a three-body interaction and a next-nearest repulsion.
The on-site trimer and trimer superfluid  exist if the on-site two-body repulsion and three-body repulsion are smaller than some thresholds.
With atom-tunneling terms, the phase transitions from trimer superfluid phase to both
atom superfluid and atom supersolid phases are first order.
With $W/J=12$, in a one dimensional lattice, the trimer superfluid phase does not exist at all.
In contrast, the trimer superfluid phase,  exists in the lower density regions $0 \textless \rho \textless2$ on
the square lattice if $J$ is not very large.
The trimer superfluid phase emerges in a wider range $0 \textless \rho \textless3$ in
the triangular lattice, or in the cubic lattice ($z=6$).
When the three-body interaction is turned on,  a trimer supersolid phase emerges due to
 the classical degeneracy between the quasi-trimer solid and the trimer solid being broken by the quantum fluctuation.
 The phase transitions from the trimer supersolid phase to the quasi trimer solid are first order and the phase
transition from the trimer supersolid phase to trimer solid  is continuous.
Our results, obtained by mean-field and quantum Monte Carlo methods, will be helpful in realizing the trimer superfluid and supersolid by cold atom experiments.

\end{abstract}

\pacs{67.85.Hj, 03.75.-b, 67.80.kb }
\maketitle
\section{Introduction}
\label{sec:intro}
So far, trimer states have been realized in cold atom experiments,  by several  alkali metal atoms,
such as   $^{7}$Li \cite{li71,li72},  $^{39}$K \cite{39K}, $^{85}$Rb \cite{85Rb},
$^{133}$Cs \cite{133Cs} and Li-Cs \cite{chengchin}.
The schemes involving magnetic Feshbach resonance (FR) \cite{li71}
and photoassociation (PA) \cite{PA,photoasso}, or combining the latter two methods \cite{photomag}
 allow us to control the binding energies of the trimers  in a wide range from $0$ Hz to several MHz.
Although the  trimers in current experiments \cite{li71,li72,39K,85Rb,133Cs,chengchin}
are just trapped in a local well within particular few-body systems, studies
can provide us some universal properties \cite{chengchin,universal,Bohuang} and
a possibility to study the correlation effects for the trimers in the optical lattice.

The strongly-correlated effects lead to  the  trimer superfluid(TSF) \cite{efimov,trimernature, zhaihui, tsf, qpt},
in which the laser-driven three-atom combinations
move together to the nearby lattice, without any
separation.
In some parameter regions, 
 the three-atom combination on a particular site
has a nonzero correlation with the holes at an infinite distance.
Consequently, the off-diagonal long-range order (ODLRO) \cite{offorder}
takes form,  i.e.,  $\lim \limits_{r\to\infty}a_i^{\dagger3}a_{i+r}^3\ne 0$.

The  Bose-Hubbard model, with trimer tunneling terms and other density-dependent tunneling terms,
has been proposed by two of the authors \cite{tsf}.
However, the nearest-neighbor repulsion, which can be realized by dipolar interaction \cite{extended},
leads to various interesting solid phases, with broken translation-symmetry \cite{bowtie}.
Furthermore, if the solid order coexists with the ODLRO simultaneously, then
the  supersolid (SS) phase emerges.
Previously, various novel SS phases were observed,
such as the pair SS (PSS) phase on the  bilayer lattice \cite{biss}, the triangular lattice with pair
hopping terms \cite{correlated1,correlated2, pairss}, in two-species systems \cite{2ss}, f-wave SS  phases \cite{fwavess}
and molecular systems \cite{molecular}.
However, it is still unknown whether or not, the trimer supersolid (TSS),
a type of  supersolid with TSF order exists in real systems.

On the other hand, there is a condition for the existence of the TSF phase in the
state-dependent lattice \cite{tsf} as the three-body interaction $W$
will compete with the kinetic energy $-zJ$.
The larger the coordination number is, the lower the kinetic energy becomes.
In the one dimensional system,
without the magnetic Feshbach resonance, the ratio
will be kept at $W/J=12$, at which the TSF phase will disappear and the Mott insulator will appear.
Although modulating $W$ independently  can be achieved by combining the FR and PA methods, this will
add to the complexity of the experiment.

Fortunately, for the two dimensional system, the $ \rm TSF$ phase still exists at the point $W/J=12$ \cite{tsf}
when using only the PA method. Therefore, it is easier to find the TSF phase in the
two dimensional system than in the one dimensional system in cold atom
experiments and consequently we will choose a square lattice at $z=4$ and a triangular
lattice at $z=6$ as the platforms by which to study the TSF and the TSS phases.

The outline of this work is as follows.
Section \ref{sec:model} shows the Hamiltonian model and  the
classical limit of the model on both a square and triangular lattice.
Section \ref{sec:method} shows the mean field (MF) and quantum Monte carlo (QMC) methods and useful observables.
We then present the ground-state phase diagram and the TSF phase
in the square lattice in Sec.~ \ref{sec:res-sq} and results for the triangular lattice
in Sec.~ \ref{sec:res-tri}  for various cases.
Concluding comments are made in Section \ref{sec:conclu}.

\section{Model, Classical limit}
\label{sec:model}
\subsection{Model}
The starting point of our study is the Hamiltonian given by:
\bea
H &=&-t \sum\limits_{\la i,j \ra} (a_{i}^{\dag }a_{j}+a_{j}^{\dag}a_{i}) 
-J\sum\limits_{\la i,j \ra} (a_{i}^{3\dag}a_{j}^3+a_{j}^{3\dag}a_{i}^3)
\nonumber \\
&&  +\frac{U}{2}\sum\limits_{i}n_{i}(n_{i}-1)+V\sum\limits_{\la i,j \ra} n_{i}n_{j}-\mu\sum\limits_{i}n_{i} \nonumber \\
&& +\sum\limits_{i}\frac{W n_i(n_i-1)(n_i-2)}{6} ~,
\label{H}
\eea
where $\la i,j \ra$  denotes the nearest-neighbor sites,
$a_i^{\dag} (a_i)$ is the boson creation (annihilation) operator at site $i$,
and $n_i=a_i^{\dag}a_i$  the boson number operator.  The term
$t$~$(J)$ is the single-atom (trimer) hopping amplitude,
$U$  the on-site interaction,  $\mu$ the chemical potential,
$V$ the nearest-neighbor repulsion, and $W$ the three-body interaction.
These parameters in the Hamiltonian can be
expressed as:
\begin{subequations} \label{eq:1}
\begin{align}
J &=  2\frac{\Omega_1\Omega_2}{\Delta} \label{J}\\
  U&=U_0-\frac{12}{\Delta}(\Omega_1^2+\Omega_2^2 ) \label{U}\\
 W&=W_0+\frac{12}{\Delta}(\Omega_1^2+\Omega_2^2 ) \label{W}\\
V&=U_0 e^{-\frac{3}{2}\pi^2\sqrt{\frac{V_0}{E_R}}} ~,
\end{align}
\end{subequations}
where  $U_0$ and $W_0$ are the traditional two-body repulsion term \cite{mean3} and
three-body repulsion term respectively, and $\Delta$ is the detuning parameter,
$V_0/E_R$ is the ratio between the depth of the optical well to the recoil energy of the atoms.
The ratio $W/J$ becomes a fixed value $12$  for any values of $\Delta$
due to $W_0$ being very weak and therefore negligible.
The coupling between atoms is defined as:
\begin{equation}
\Omega_{1(2)}=\Omega\int ~d\mathbf{r} ~ w_m^{*}( \mathbf {r} - \mathbf { r_{i_m}} )w_a(  \mathbf {r} -
\mathbf {r_{i_a(i_a+1)})}^3 ~,
\end{equation}
where $w_{a(m)}(x-x_i)$  are the ground state Wannier functions for an atom (a trimer) in an optical lattice potential.
If $t$ or $J$  dominates the Hamiltonian, the system sits in the atom superfluid (ASF) or TSF phase.
If the remaining  interactions  dominate
the Hamiltonian, the system will enter into the solid  phases or Mott insulator (MI) phases.
In some regions, we also expect a TSS phase.

\begin{figure}[htb]
\includegraphics[width=0.95 \columnwidth]{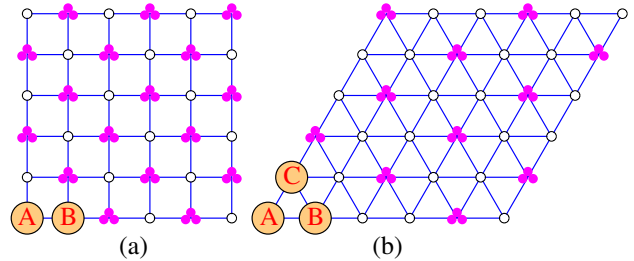}
\caption{(Color online) (a) Trimer solid phase ($0$, $3$) on the square lattice,
where the numbers ($n_A$, $n_B$) denote the boson occupation numbers
in the sublattices. The three magenta symbols and empty symbols denote
trimers and empty sites, respectively.
(b) Trimer solid phase $(0, 0, 3)$  on the triangular lattice.
The trimer solid phase $(0, 3, 3)$ and other quasi-trimer solids such as $(0, 1, 3)$ and $(2, 3, 3)$ are not shown.}
\label{snap}
\end{figure}

\subsection{Classical limit: trimer solid}
\label{sec:t0J0}

This section shows that the trimers still exist if
the two-body and three-body interactions are
in reasonable ranges.

In the classical limit,  $t= 0$, $J= 0$ and the nearest neighborhood
repulsion $V$ in the Hamiltonian of $\rm Eq.$~(\ref{H})  breaks translational symmetries
of the original lattice.  A minimum  cell in the solid phases is composed
of $ \rm two$ (three) sublattices  $A$ and $B$ ($A$, $B$ and $C$)  for a square (triangular) lattice
as shown in Fig.~\ref{snap}.
We use $n_A$, $n_B$, and $n_C$ to denote the occupation numbers, and the Hamiltonian in a  cell
becomes:
\be
H     = \sum_{i}  H_i + H_V ~,
\ee
where the on-site Hamiltonian is:
\[
H_i   = \frac{U n_i ( n_i - 1 )}{2} +  \frac{W n_i( n_i - 1 ) (n_i - 2 )}{6}-\mu n_i ~,
\]
and the sum is taken over the sites in the cell. Here, the interaction term  is:
$H_V = \frac{zV}{2} ( n_A n_B + n_B n_A )$ for the square lattice and:
\[
H_V = \frac{zV}{2} ( n_A n_B + n_B n_C + n_C n_A)
\]
for the triangular lattice.
The coordination number $z$ is $4$ for the square lattice  and $6$ for triangular lattice.

\begin{figure}[htb]
\includegraphics[width=\columnwidth]{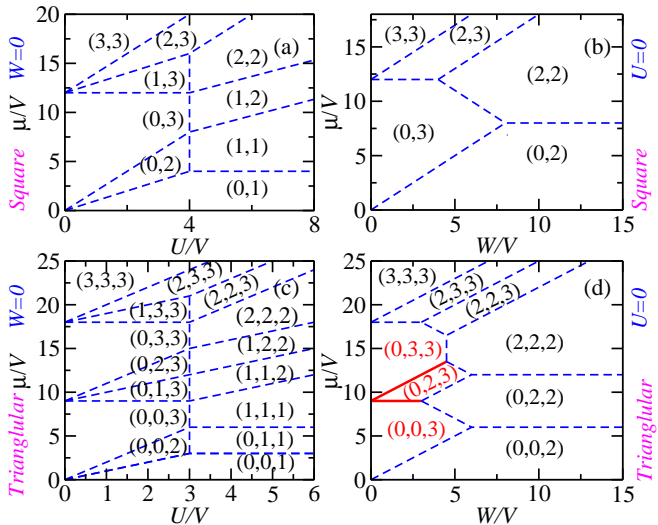}
\caption{(Color online) Classical limit ($t=0$, $J=0$)  of the BH model with (left)
and without (right) the three-body interaction.
The upper row is for the square lattice and the lower row is for the triangular lattice.
(d) The trimer solids  (0, 3, 3) and (0, 0, 3) are separated with $W/V\ne0$ while the two solids are degenerate with
the quasi solid (0, 2, 3) along the solid lines.}
\label{t0J0}
\end{figure}

We can obtain the phase boundaries between various ordered phases
by comparing the energies per unit cell in the phases.
The phase diagrams of the square lattice and  the triangular lattice  are  shown
in the upper and lower row respectively.
We show the phase diagram
in the plane  ($U/V$, $\mu/V$) with $W=0$ in Figs.~\ref{t0J0} (a) and (c),  and the phase diagram   ($W/V$, $\mu/V$) with $U=0$
in Figs.~\ref{t0J0} (b) and (d).

In the square lattice, as shown in  Fig.~\ref{t0J0}(a),
only one solid phase $(0,~3)$ exists with $U=0$.
Due to the breaking of the trimer by the nonzero two-body interaction,
quasi trimer solid phases ($1$, $3$) and ($2$, $3$) emerge as $U/V \textless 4$.
The remaining on-site trimers are broken and solid phases
($0$, $1$) and ($1$, $2$) exist without trimers if $U/V \geq 4$.
In the limit $U \rightarrow \infty$, the system reduces to
hardcore bosons, and only two possible occupation numbers  $0$ and $1$  are allowed, which are not shown.
Similarly, in Fig.~\ref{t0J0}(b), the maximum range of the trimer  solid $(0, 3)$  is  $W/V	\textless 7.5$.

Figs.~\ref{t0J0}(c) and (d)  show the effects of $U$ and  $W$  on  the trimers in the triangular lattice.
The trimer solids  ($0$, $3$, $3$),  ($0$, $0$, $3$) and the quasi trimer solid
( 0, 2, 3), which are of special interest, exist in the range $U/V\textless 3$ and $W/V\textless 5$
as shown in Figs.~\ref{t0J0} (c) and (d).
With $U =0$ and $W=0$,  the system  resembles the  hardcore BH model on the triangular lattice, just by replacing the solids  (0, 0, 3)  and (0, 3, 3) with (0, 0, 1) and (0, 1, 1) respectively.
Due to the geometrical frustration, in the classical limit, the solids (0, 0, 1) and (0, 1, 1) are degenerate
at the hole-particle symmetric point  and the system has a perfect solid order with the wave vector $\mathbf{Q}=(4\pi/3,0)$ \cite{mechanism,mecha1}. Turning on the hopping terms,   the solid orders would be  partially  broken by the quantum fluctuation with the SS order forms coexisting   with the solid orders.

As shown in  Fig.~\ref{t0J0}(d),  the two solids ($0, 0, 3$) and ($0, 3, 3$) are  not degenerate and are separated  by the solid ($0, 2, 3$)  in the phase diagram with $W/V\ne 0$.  One may infer that the supersolid does not exist as the degeneracy is broken.
However, the quasi trimer solid (0, 2, 3) is degenerate with the solids ($0, 0, 3$) and ($0, 3 ,3$) along the solid lines in  Fig.~\ref{t0J0}(d). With trimer hopping,  it is necessary to check  the possibility  for  the
existence of the TSS phase, which will be discussed in Sec.~\ref{sec:triw12}.

\section{methods and the measured quantities}
\label{sec:method}
\subsection{Mean-field method}
\label{sec:meanfield}
By the decoupling approximation \cite{wzg,dec,TMF2} defined by:
\be
O_{i}^{\dag}O_{j}=\la O_{i}^{\dag}\ra O_{j}+O_{i}^{\dag}\la O_{j}\ra - \la O_{i}^{\dag}\ra \la O_{j}\ra
\ee
we can transform the model Eq.~(\ref{H}) into the mean-field form by
decoupling the terms $a_{i}^{\dag}a_{j}$, $a_{i}^{3\dag}a_{j}^3$ and $n_{i}n_{j}$ into single site forms.
The Hamiltonian on the triangular lattice can be considered as the sum over local
Hamiltonian operators on the three sublattices $ A $, $ B $, and $ C $.
The Hamiltonian on the sublattice $ A $, $ h_A $, is as follows:
\bea
h_A&=&-3t (a^{\dag}_{A}+ a_{A})(\Psi^a_B+\Psi^a_C)+ 3t\Psi^a_A(\Psi^a_B+\Psi^a_C)\nonumber\\&&
-3J (a^{3\dag}_{A} + a^3_{A})(\Psi^t_B+\Psi^t_C)+ 3J\Psi^t_A(\Psi^t_B+\Psi^t_C) \nonumber\\
&& +3Vn_A(\rho_B+\rho_C)-3V\rho_A(\rho_B+\rho_C)/2\nonumber\\\label{eq:hamt}
&&+U[n_A(n_A-1)]/2-\mu n_A\\&&
+W n_A(n_A-1)(n_A-2)/6 ~, \nonumber
\eea
where $\Psi^a=\la a \ra$,$\Psi^t=\la a^3 \ra$, and $\rho=\la n \ra$.
For the square lattice, we only need to replace the above Hamiltonian with the following:
\bea
\hat h_A &= &-4J\Psi_{B}^{t}(\hat a_{A}^{3}+\hat a_{A}^{\dagger 3})+\hat n_{A}(-\mu+ 4\rho_{B}V)\nonumber\\
&&+4J\Psi_{A}^{t}\Psi_{B}^{t}-2\rho_{A}\rho_{B}V \nonumber \\
&&+U[n_A(n_A-1)]/2-\mu n_A\label{eq:hams} \\
&&+W n_A(n_A-1)(n_A-2)/6 ~. \nonumber
\eea
One can obtain the  Hamiltonian operators  $h_B$ and $h_C$ by  cyclic permutation of the subscripts $A$, $B$, and $C$.

We resolve the Hamiltonian operators and obtain the solutions of order parameters  iteratively.
The expectation values of the creation (annihilation) operators are considered
as variational parameters in a manner similar to that of references \cite{kapa,mean2, mfde}.

One can initiate a multiple-group of trial parameters $(\Psi^a_j,\Psi^t_j,\rho_j)$,
where $j$ = $A$, $B$ and $C$, and then solve the eigenvalue equation of $h_A$.
We input the newly calculated quantities  $\Psi^a_A$, $\Psi^t_A$, and $\rho_A$  into  $h_B$
and $h_C$ sequentially. We iterate until the order parameters converge. In order to filter out the
metastable states, we save the states of  the lowest ground-state energy, by which
we get the ground-state phase diagrams.

In Tab. \ref{Tab:t1}, we show the values of the order parameters for several typical phases.
The solid orders  
denoted by  $\Delta \rho$, $\Delta \Psi^a$ and $\Delta \Psi^t$ are defined by:
\be \label{eq:2}
\Delta O  = \sum_{i=A,B,C} |O_i-\bar{O}|,~~~~ \bar{O}=\frac{1}{3}\sum_{i=A,B,C}\bar O_i,
\ee
where $O$ can be replaced by $\rho$, $\Psi^a$, and  $\Psi^t$.
More details are
available in the work of references \cite{wzg,pairss,tsf}.
 \begin{table}[t]
\caption{
Values of the order parameters for typical phases.
}
\begin{center}

\begin{tabular}{c|cccccc}
\hline
\hline
          ~     & ~ ~ASF~~ &~~Solid~~  &  ~~ASS~~    & ~~ TSF~~ &~~TSS~~  & ~MI~    \\ \hline
 $\Psi^a$       &  $\ne0 $ &   0       & $\ne0$      &     0    &  0      &   0          \\
 $\Psi^t$       &  $\ne0 $ &   0       & $\ne0$      & $\ne0$   & $\ne0$  &   0          \\
 $\Delta\rho$   &    0     & $\ne0$    & $\ne 0$      &    0     & $\ne0$  &   0          \\
$\Delta\Psi^a$  &    0     & $\ne0$    & $\ne0$    &  0         &   0      & 0        \\
$\Delta\Psi^t_t$&    0     & $\ne0$    & $\ne0$    &  0         & $\ne0$   & 0        \\\hline
$\rho_s^{a}$    &   $\ne0$    & $0$    & $\ne0$    &  0         & $0$   & 0        \\
$\rho_s^{t}$    &    $\ne0$     & 0    & $\ne0$    &  $\ne0$        & $\ne0$   & 0        \\
$S({\bf{Q}})/N$    &    0     & $\ne0$    & $\ne0$    &  0         & $\ne0$   & 0        \\
\hline
\hline
\end{tabular}
\label{Tab:t1}
\end{center}
\end{table}


\subsection{Improved Quantum Monte Carlo method}
To check the results obtained from the MF method,
we simulate the model (\ref{H})
using the stochastic series expansion (SSE) QMC method\cite{sse}
with the directed loop update\cite{directedloop}.
To simulate the pair superfluid phase,
the algorithm is improved by allowing the
head of a directed loop to carry a pair \cite{lode,psf,mfy2, fintemp}  of creation (annihilation) operators
$a^{2\dag}$($a^{2}$). In the present work, to make the atom trimers active, we let the loop heads carry
trimer creation (annihilation) operators  $a^{3\dag}$($a^{3}$).

To distinguish the ASF and the TSF states,
we define two types of  superfluid stiffness
$\rho_s^{\alpha}$ as the order
parameter\cite{sfs,tsf}, which can be determined as:

\be
\rho_s^{\alpha}=\frac{1}{2} \left(\rho_s^{\alpha,x}+\rho_s^{\alpha,y}\right),
\ee
where $\rho_s^{\alpha,x}$ and  $\rho_s^{\alpha,y}$ are the superfluid stiffness along  $x$ an $y$ directions respectively.
In the above equation,  $\rho_{s}^{\alpha,x}$ is defined as
\begin{equation}
\rho_{s}^{\alpha,x}=\frac{L^{2-d}\langle W(\alpha,x)^2\rangle}{2d\beta(9J+t)}.
\label{rho}
\end{equation}
If $\alpha=t$,
$W(\alpha,x)$ is the winding number which can be divided without remainders.
If $\alpha=a$,
$W(\alpha,x)$ is the winding number which cannot  be divided without remainders.
The  definition  of $\rho_{s}^{\alpha,y}$ is  similar with that of  $\rho_{s}^{\alpha,x}$.
For a TSF phase,  we define the  order parameter $\rho_s^{t}>0$ and $\rho_s^{a}=0$.
For an ASF phase, $\rho_s^{a}>0$ and  $\rho_s^{t}=0$.
The parameters $d$ and $L$ are respectively the system dimensionality and size, and $\beta$ is the inverse temperature.

The structure factor is defined to characterize the solid  order:
\begin{equation}
S({\mathbf Q})/N=\langle \rho_{{\mathbf Q}} \rho^{\dagger}_{{\mathbf Q}} \rangle,
\end{equation}
where
$\rho_{{\mathbf Q}}=(1/N)\sum_i n_i \exp(i {\mathbf Q} {\mathbf r_i})$.
The trimer solids (0, 0, 3) and (0, 3, 3) share the same ordering at
the wave vector ${\mathbf Q}=(4\pi/3, 0)$ and the same value 1 in the
perfect ordering. In the square lattice, the wave vector
should be  ${\mathbf Q}=(\pi,\pi)$ for the solid (0, 3).
\begin{figure}[htb]
\includegraphics[width=\columnwidth]{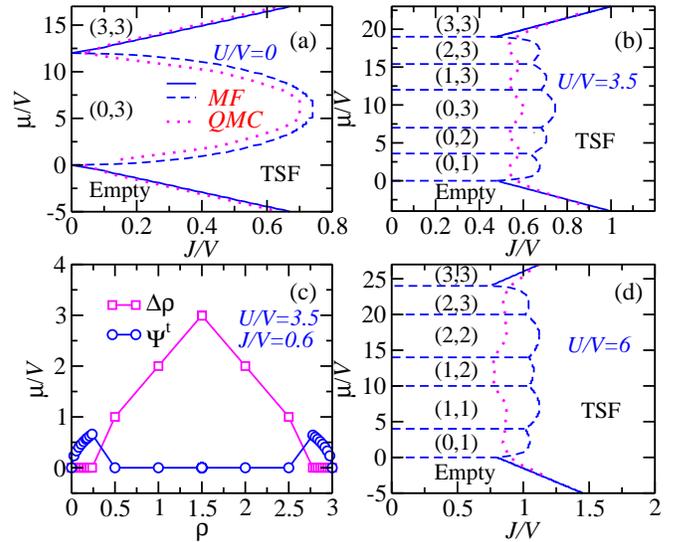}
\caption{(Color online)
MF and QMC  ground-state phase diagrams for $U/V=0$ (a) , $3.5$ (b), and $6$ (d), respectively.
Solid-dashed  lines by MF method  denote first-order phase transitions,
whereas solid lines represent the continuous phase transitions.
The doted lines are obtained by the QMC method.
(c) The hole-particle symmetry is shown by plotting
$\Delta \rho$ and $\Psi^t$ as function of $\rho$.
}
\label{w0u046sq}
\end{figure}

\section{ASS and TSF on the square lattice}
\label{sec:res-sq}
\subsection{Solid and TSF with $t/V=0$ and $W/V=0$ }
The  phase diagrams  for  $U=0$, $3.5$, and $6$,
are shown in Figs.~\ref{w0u046sq} (a), (b) and (d), respectively.
Several typical trimer solid phases are found. Also, with larger
$J/V$, a TSF phase emerges.
The phase transition from the solid phase to the TSF phase is first order and
the phase transitions from
the empty phase to the TSF (MI) phase are continuous.
No supersolid phase is found, which is consistent with
the hardcore bosons in  biparite  lattice\cite{separation,rmpSS}.
These results are obtained by solving the equations described in section \ref{sec:meanfield}.

To understand the phase diagram, we also try to solve the MF Hamiltonian analytically.
In the TSF phase, there is no symmetry broken between
the two sublattices $A$ and $B$ and these two sublattices are equivalent, so we can set  $\Psi_A^t=\Psi_B^t$ and
$\rho_A=\rho_B$ in Eq.~(\ref{eq:hams}). In this way, the number of the variables will be reduced.
By solutions for the minimum value i.e. the lowest eigenvalue of the Hamiltonian in Eq.~(\ref{eq:hams}), we find that
the density is:
\be
\rho=\frac{24J +3\mu}{16J + 12V},
\label{rhosq}
\ee
and the trimer superfluid order parameter is:
\be
\Psi^t=\frac{\sqrt{3(8J + \mu)(8J + 12V - \mu)}}{2\sqrt{2}(4J + 3V)}.
\label{psisq}
\ee
From Eqs.~(\ref{rhosq}) and (\ref{psisq}), we can obtain the phase boundaries  between the TSF phase and  the solid(MI) phases.
Letting $\Psi^t=0$ in Eq.~(\ref{psisq}),  we get the  straight  phase boundary
$ \mu/ V=-8J/V$ from the  TSF phase to the empty phase, and
the straight phase boundary $\mu/ V=8J/V+12$ from the TSF phase to the MI phase.
This can also be understood by  the single particle theory \cite{tsf}.

The boundary line from the solid ($0$, $3$) to the TSF phase, can also be obtained.
By substituting  Eq.~(\ref{rhosq}) and Eq.~(\ref{psisq}) into the lowest eigenvalue,  we get:
\be
E=- (24J + 3\mu)^2/(16J + 12V) ~.
\ee
By letting $E=-1.5\mu$, where the right side of the previous equation
is the energy per site in the solid ($0$, $3$).
The boundary line  of the  solid phase  to the TSF phase is obtained as follows:
\be
\mu/V=6\pm 2\sqrt{ 9-16J^2/V^2},~ (J/V\le 3/4).
\ee
The equation above is consistent with the numerical solutions.
The triangular symbols in the phase diagram  are the data obtained by
QMC method. The position of the symbols is consistent with the boundary lines
by the MF method.

With any nonzero value of $U$,
phase transition boundary lines can also be obtained analytically.
For example,
the boundaries between the solid ($0,1$) to the TSF phase are:
\begin{equation}
\mu=C \pm \frac{12V+16J}{3} \sqrt{\frac{3U - 20J + 3V}{16J + 12V}} ~,
\end{equation}
where $C= U+2V-16J/3$.

In Fig.~\ref{w0u046sq} (c),
we scan the phase diagram $U/V=3.5$ along $J/V=0.6$ and find that the above phase diagram is particle-hole symmetric by the
transformation  $\rho\rightarrow 3-\rho$. In the region $0\textless\rho\textless0.25$ and
$2.75	\textless\rho	\textless3$, the system is in the TSF phase.
In the range $0.25\textless\rho	\textless2.75$, the density changes discontinuously, which means the phase
transition between the TSF phase and the solid phases,
 and phase transitions between solid phases, are first order.

\begin{figure}[htb]
\includegraphics[width=\columnwidth]{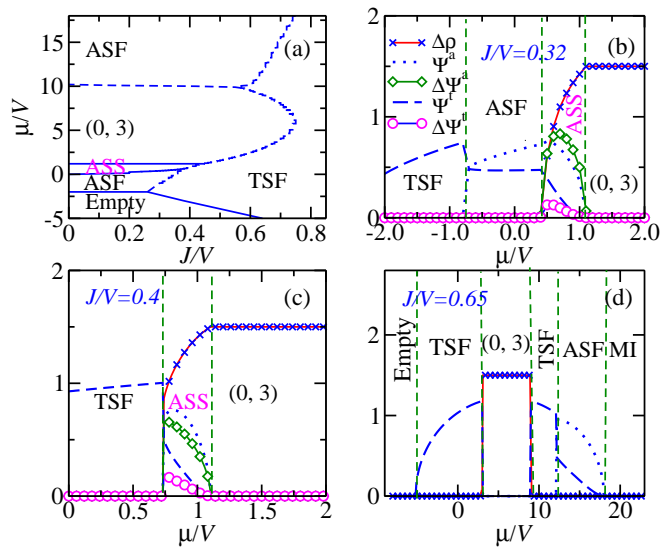}
\caption{(Color online) (a) MF phase diagram of the model  represented
 by Eq.~(\ref{eq:hams}) for $U/V=0$, $t/V=0.5$, and $W/V=0$. The dashed line
and the solid line denote the first-order and the continuous phase transitions, respectively.
Detailed description of $\Delta \rho$, $\rho$, $\Delta \Psi_t$, and $\Psi_t$ for
the square lattice  with $U=0$, $t=0$ and $W=0$ at  (b)$J/V=0.32$  (c) $J/V=0.4$ and (d)$J/V=0.65$. }
\label{t0.5}
\end{figure}

\subsection{ASS with $t/V\ne 0$, $W/V=0$ and $U/V=0$}
In Fig.~\ref{t0.5}(a), with a moderate atom hopping $t/V$, i. e., $ t/V=0.5 $,
two phases namely the ASF and ASS phases emerge between the empty phase and the solid phase $(0, 3)$.
The phase transition from the ASF phase to the TSF phase is of first order
due to the quantum  fluctuations \cite{tsf}. The phase transition
from the ASS to the TSF phase is also first order.

To illustrate the results above in more detail, we $\rm scan$
the phase diagram along the lines  $J/V=0.32$, $0.4$, and $0.65$,
as shown in Figs.~\ref{t0.5} (b)-(d), respectively. With $J/V=0.32$ and starting at $\mu/V=-2$,
the system sits in a TSF phase with  $\Psi^t \ne  0 $ and  $\Psi^a =0$.
When increasing the chemical potential $\mu$, the system enters into the ASF phase discontinuously.
The order parameter $\Psi^{a}$
jumps to a nonzero value, and $\Psi^t$ deceases steeply. The jump means the ASF-TSF
phase transition is first order, which is consistent with
our previous work  \cite {tsf}. In Fig.~\ref{t0.5}(c), the phase
transition from the TSF to the ASS phase is first order while the
phase transition from  the ASF to the ASS phase is continuous as shown in Fig~\ref{t0.5} (b).
In Fig.~\ref{t0.5}(d), with $J/V=0.65$, the ASS phase disappears and the TSF and ASF phases dominate most of the phase diagram.

\subsection{TSF with $t/V=0$, $W/J =12$ and $U/V=0$}

In order to make the real ultra cold atom experiment easier to
observe the TSF phase, we try to make
the PA method  drive the trimers active in
the optical lattice.
We don't try to combine the PA method with the magnetic Feshbach resonance method together
to modulate the three-body interaction independently.
The price of just using the PA method
 is that the ratio $W/J$  has to be kept at $12$
and is not controllable.

Fortunately, we find in Fig.~\ref{w12ju0sq} (a) the TSF phase with $W/J=12$.
In the larger $\mu$ region, the phase boundaries between the phases ($3$, $3$), ($0$, $3$),
 ($0$, $2$) and ($3$, $2$) are completely identical to the classical limit phase diagram in Fig.~\ref{t0J0}(b).
In the lower $\mu$ region, the expected TSF phase appears only in the region ($\rho \textless 2$),
in spite of the block from the competition from the nearest neighborhood
interaction and the local three-body repulsive interaction.
The numerical phase boundary between the $(0, 2)$ solid to the TSF phase are very consistent with the analytical results:

\begin{figure}[tb]
\includegraphics[width=\columnwidth]{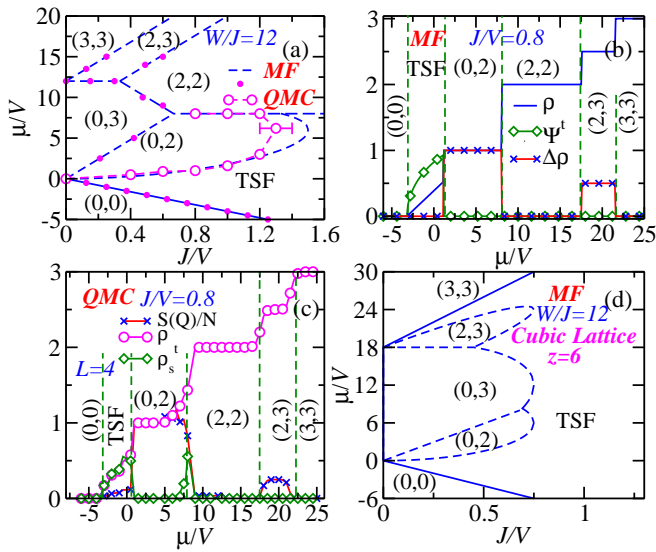}
\caption{ (Color online) (a)~MF and QMC  phase diagram ($J/V$, $\mu/V$) with $W/J=12$.
(b) By the MF method, detailed description of  the density $\rho$, the TSF order $\Psi^t$ and   solid orders  $\Delta \rho$ and  $\Delta \Psi^t$   for
the square lattice  with $U=0$, $t=0$ and $W=12J$ at  $J/V=0.8$. (c) By the QMC method, the density $\rho$, the structural factor $S({\bf Q})/N$,  and the TSF density  $\rho_s^t$ in the same parameters regimes  as in (b).
(d) The MF phase diagram for the cubic lattice with $z=6$ for comparison purposes.
}
\label{w12ju0sq}
\end{figure}
\begin{equation}
\mu=\frac{1}{6}\left(\left(8J+24V\right) \pm \sqrt{\left(8J+24V \right)^2-576J^2}\right) ~.
\end{equation}
At the same time, in the classical limit, the configuration (0, 3) disappears if $W/V >7.5$. However,
in the quantum phase diagram, when $W/V > 7.5$  the TSF phase exists.
This implies that the TSF phase benefits from the relative strength of the trimer tunneling
terms and  the on-site repulsion is not the only interaction which determines the trimers in the TSF phase.

It should be pointed out that for the  square
lattice, theoretically, the TSF phase only exists in the range $W/J \textless 24$ with just
two nonzero parameters $W$ and $J$. Experimentally, it is possible to find the $\rm TSF$ phase as the
ratio $W/J=12$ is smaller than $24$ for arbitrary detuning parameter $\Delta$.

We describe the phase diagram in more detail along $J/V=0.8$ as shown in Fig.~\ref{w12ju0sq}(b).
The system undergoes a phase transition from the empty phase to the TSF phase
at $\mu/V=-3.1$. At $\mu/V=1.1$, the TSF phase disappears and the system enters into
the $(0,2)$ solid phase with a first-order phase transition due to the jumps of the
quantities $\Psi^t$ and $\rho$.
When increasing the ratio $\mu/V$, the solid phases ($2,~2$), ($2,~3$), and the MI phase ($3$,~$3$) appear  successively.
By the QMC method, Fig.~\ref{w12ju0sq}(c) shows the details along $J/V=0.8$.
The regions are very consistent with the results by the MF method as shown in Fig.~\ref{w12ju0sq}(b).
In the region around $\mu/V=9$, $\rho_s^t$ is nonzero, but becomes zero in
the thermaldynamical limit.

In Fig.~\ref{w12ju0sq}(d),
for the purpose of comparison, we compute the quantum phase diagram
on the cubic optical lattice with coordination number  $z=6$.
Without the next-nearest repulsion, the quantum phase diagram of the Bose-Hubbard
on the lattices with different dimensions are similar.
However, with off-site interactions, the phase diagrams need to be checked, as the coordination number  $z=6$ is larger than
the square lattice, it is easier for the kinetic-dominated TSF phase to emerge.
Eventually, in contrast to the square lattice, we find the TSF phase in a wide range.

\section{TSS on the triangular lattice}
\label{sec:res-tri}
\subsection{TSS with $U/V=0$, $t/V=0$ and $W/J=0$}
\begin{figure}[t]
\includegraphics[width=\columnwidth]{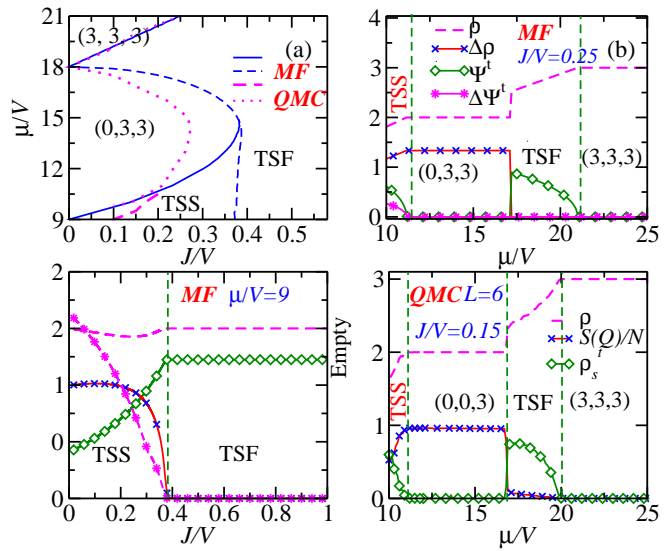}
\caption{ (Color online) (a) MF and QMC  phase diagram of the model represented by Eq.~(\ref{eq:hamt}) for $U=0$, $t=0$ and $W=0$ for the triangular  lattice.
The dashed line and the solid line denote the first order and the continuous transitions, respectively.
(b) By the MF method, detailed description of the density $\rho$,
the TSF order $\Psi^t$, and the solid orders $\Delta \rho$ and  $\Delta \Psi^t$,   with $J/V=0.25$.
(c) The same quantities as in (b) at $\mu/V=9$. (d) The density $\rho$, the structural factor $S({\bf Q})/N$,  and the superfluid density  $\rho_s^t$
with $J/V=0.15$ by the QMC method and the results are consistent with those in (b).}
\label{t0u0w0}
\end{figure}

\begin{figure}[hb]
\vskip 0.5 cm
\includegraphics[width=\columnwidth]{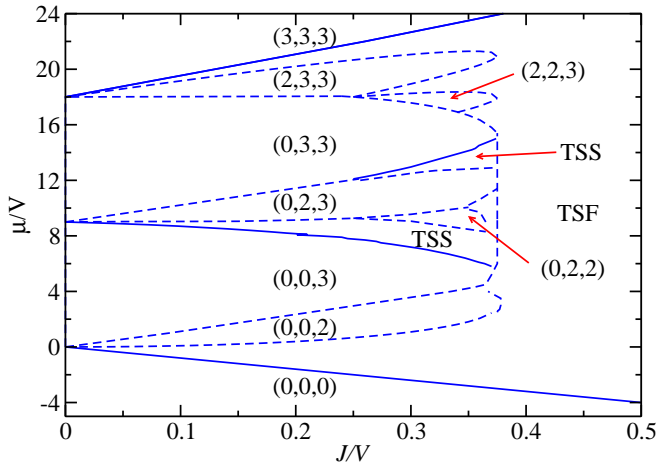}
\caption{(Color online)~MF  phase diagram of the triangular lattice for $U=0$, $t=0$ and $W/J=12$.  The dashed line and the
solid line denote the first order and the continuous transitions, respectively.
The TSS phases exists in the two independent regions. In other words, two solid
phases emerge in the original region of the TSS phase in Fig.~\ref{t0u0w0}(a).}
\label{tri}
\end{figure}

This section shows the phase diagram for  $U=0$  on the triangular lattice.
In Fig.~\ref{t0u0w0}(a), we show the phase diagram of the model described by  $\rm Eq.$~(\ref{eq:hamt}) on the triangular lattice,
which contains the empty, solid, TSF and MI phases.
The new emerged phase is the TSS phase.

In the TSF phase, $\rho$ and  $\Psi^t$ for every site are the same.
By obtaining the lowest eigenvalue of Eq.~(\ref{eq:hamt}),
the trimer superfluid order parameter becomes:
\be
\Psi^{t}=\frac{\sqrt{(12J+\mu)(12J+18V-\mu)}}{\sqrt{6}(4J+3V)}.
\ee
The phase boundaries between the TSF phase and the solid (MI) phases can also be obtained analytically; these are
not shown here.

We show the details of the phase diagram along the line $J/V=0.25$.
By both the nonzero values of the solid orders
$\Delta \rho$ and  $\Delta \Psi^{t}$, and the TSF order $\Psi^t$,
a stable TSS phase
exists between the two solid phases, i.e.: in the range $ 10 \textless \mu/V \textless 11.4$.
The mechanism is order by disorder \cite{mechanism}, in which quantum fluctuations break
the partially classical degeneracy.
At the same time, the phase transition from the TSF phase to the solid phases
are first order.

As shown in Fig.~\ref{t0u0w0} (c), we scan the phase diagram
along $\mu/V=9$. In the region $0\textless J/V \textless 0.39$,
the solid order and TSF order are robust and
coexisting.
The phase transition from the TSS to the TSF phases
are continuous in the particle-hole symmetric point $\mu/V=9$\cite{yamamoto2,swessel}.
As shown in Fig.~\ref{t0u0w0} (d), we scan the phase diagram at $J/V=0.15$ by the QMC method.
The result is again very consistent with Fig.~\ref{t0u0w0} (b) by the MF method.

We also study the effects of the on-site interaction $U/V$, not shown. We start at the point
$\mu/V=8$ and $J/V=0.25$ which is at the TSS phase.
By increasing $U/V$, the TSS phase survives in
a reasonable range and enters into the (0, 0, 3) phase.

\subsection{TSS with $t/V=0$, $U/V= 0$ and $W/J=12$}
\label{sec:triw12}

\begin{figure}[htb]
\includegraphics[width=0.95 \columnwidth]{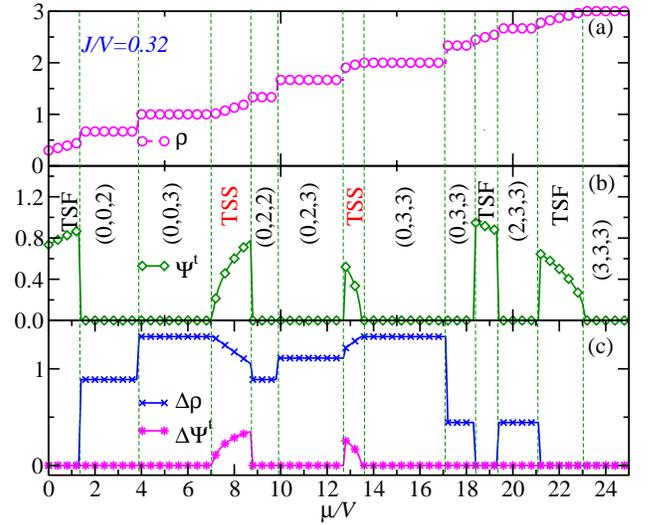}
\caption{(Color online)
MF detailed description of the density $\rho$,  the TSF order $\Psi^t$ and  solid orders  $\Delta \rho$ and  $\Delta \Psi^t$, for the triangular  lattice with $U=0$, $t=0$, $W/J=12$ and  $J/V=0.32$. The TSS phases exist in the two regions $7.1\textless\mu/V\textless8.7$ and $12.8\textless\mu/V \textless13.4$.}
\label{d}
\end{figure}

Figures~\ref{tri} shows the global phase diagram for the triangular lattice with $W/J=12$.
When the ratio $J/V$ is relatively weak, several kinds of solid phases are found as shown in the classical limit phase
diagram Fig.~\ref{t0J0}(d).
As the ratio $J/V$ become stronger, the expected TSF phase appears.
In this phase diagram, the TSF phase can be found in the range $0\textless \rho \textless 3$, which is significantly
different from the case on the square lattice.
All of the phase boundaries between the TSF phase and the solid phase obtained by the iteration method have been checked by
analytical evaluation.

Fortunately, apart from the previous phases, we find two independent
regions of the TSS phases.
One may think this is an unexpected result because, in the classical limit
phase diagram, the two types of solid (0, 0, 3) and
(0, 3, 3) are not degenerate, which is the condition that the
supersolid exists \cite{mechanism,mecha1,separation,rmpSS}.
However,  the quasi trimer solid (0, 2, 3) is degenerate with the trimer solid (0, 0, 3).

Comparing with Fig.~\ref{t0u0w0}(a),
the areas of the TSS phase become smaller and the solids (0, 2, 3)
and (0, 2, 2) appear between the two regions of the TSS phases.
Moreover, the phases (0, 0, 2), (2, 3, 3) and (2, 2, 3)
emerge due to the break of the trimers by the three-body repulsion.

Figure~\ref{d} shows the  detailed descriptions of the density $\rho$, the TSF order
 $\Psi^t$ and  solid orders  $\Delta \rho$ and  $\Delta \Psi^t$
along $J/V=0.32$. The TSS phase exists in the range $7.1\textless\mu/V\textless8.7$
 and $12.8\textless\mu/V \textless13.4$, for both $\Psi_t$ and $\Delta \rho$
are nonzero.
Also there are several solid phases, in which $\Delta \rho$ and $\rho$ are in some horizontal platforms.
Furthermore, two TSF phases appear in the three independent regions.
 Unlike the square lattice case, the density of the TSF phase covers the range $0\textless \rho \textless 3$.
The phase transitions from the TSF phase to the solid phases (0, 0, 3) and (0, 3, 3) are continuous,
 while the phase transitions from the TSF phase to the quasi trimer solid
 ($0$, $2$, $3$) and the other phase  ($0$, $2$, $2$)  are first order.

\begin{figure}[htb]
\includegraphics[width=0.95 \columnwidth]{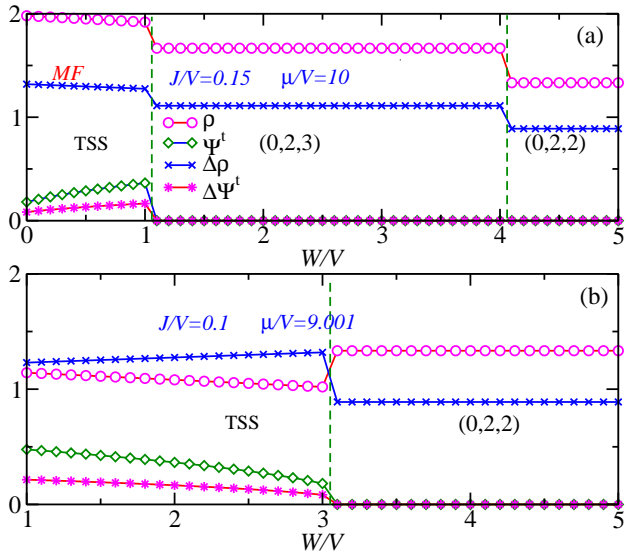}
\caption{(Color online)
MF description of the density $\rho$, the TSF order $\Psi^t$, and solid orders $\Delta \rho$
and $\Delta \Psi^t$ as functions of $W/V$ at  (a) $\mu/V=10$ and  $J/V=0.15$, and (b) $\mu/V=9.001$ and $J/V=0.1$.
The jumps of the order parameters confirm that the phase transition from  TSS to the quasi trimer solid phase
is first order.}
\label{last}
\end{figure}

To investigate the phase transition from the TSS phase to the quasi trimer solid phase,
we also start a TSS phase, where has been proved in Fig.~\ref{t0u0w0}(d),  then increase the three-body interaction.
Fig.~\ref{last} shows the quantities $\rho$, $\Delta \rho$, $\Psi^t$ and $\Delta \Psi^t$ as functions of
$W$ at   $\mu/V=10$ and $J/V=0.15$. From the jumps of these quantities around $W/V=1$, the phase transition
from the TSS  phase to the solid  phase  (0, 2, 3) clearly is first order.
We also find  that the transition from the TSS to the solid (0, 2, 2) is first order at $J/V=0.1$  and $\mu/V=9.001$ around
$W/V=3.05$.

\section{discussion and conclusions}
\label{sec:conclu}

Most of results in the present work were obtained by the MF method.
This method ignores quantum fluctuation terms and correlation effects
by a decoupling approximation \cite{dec,TMF2},
but was nonetheless successful at predicting  SS phases \cite{iskin, yamamoto2} and
phase transitions \cite{yamamoto1, takashi},
that were eventually confirmed by
QMC calculations \cite{qmc, swessel} or the density matrix renormalization group method \cite{xfzhang}.
In this work, we checked our MF results by the QMC  method in some applicable regions and found that the results from both methods
were very consistent.

In conclusion,
we obtain the phase diagram of the extended Bose-Hubbard model with explicit trimer tunneling terms and a three-body interaction, on both the square and
the triangular lattices.

For the square lattice, with a fixed two-body interaction,
a TSF exists if the trimer tunneling terms are large enough.
With finite atom hopping, an ASS phase exists,
and the phase transition between the ASS phase and
TSF phase is of first order.
With a nonzero three-body interaction $W/J=12$, the
TSF phase {\relax exists} with the $0\textless \rho \textless 2$ if $J$ is not sufficiently large.
For comparison, we calculated the
cubic lattice with $z=6$, and
the TSF phase emerges in the larger range $0\textless \rho \textless 3$.

For the triangular lattice, without the three-body
interaction,  we find a stable TSF phase and
also the TSS phase.
In particular, with the three-body repulsive interaction, we also find a
stable TSS phase, due to
the classical degeneracy between the quasi trimer solid and the trimer solid being broken by the quantum fluctuation.
The phase transitions from the TSS phase to the quasi trimer solid are first order and the phase
transition from the TSS phase to the trimer solid becomes continuous.
Our results will be helpful in realizing the TSF and TSS phases by cold atom experiments.

\acknowledgments
W. Zhang thanks Songsong Wang  in checking some numerical results. W. Zhang  is supported  by the NSFC under Grant
No.11305113, No.11204204, Foundation of Taiyuan University of Technology 1205-04020102.
T. C. Scott is supported in China by the project GDW201400042 for the ``high end foreign experts project''.
C. Ding is supported by the NSFC under Grant No.11205005, Foundation of Anhui Province
1408085MA19.

\appendix

\end{document}